\documentclass[prl,a4paper,twocolumn,superscriptaddress,showkeys]{revtex4}
\pdfoutput=1
\usepackage[version=3]{mhchem}
\usepackage{graphicx}
\usepackage[english]{babel}
\usepackage{array}
\usepackage[T1]{fontenc}
\usepackage[usenames,dvipsnames]{xcolor}
\usepackage{amssymb}
\usepackage{siunitx}
\usepackage{hyperref}
\usepackage{xspace}

\hypersetup{pdfauthor={Hassan Srour, Martien Duvall Deffo Ayagou, Thi Thanh-Tam Nguyen, Nicolas Taberlet, Sébastien Manneville, Chantal Andraud, Cyrille Monnereau and Mathieu Leocmach},pdftitle={Ion pairing controls rheological properties of ``processionary'' polyelectrolyte hydrogels}}

\newcommand{\figSsec}{S1\xspace}
\newcommand{\figSmassPImBr}{S2\xspace}
\newcommand{\figSmassPPyBr}{S3\xspace}
\newcommand{\figSmassPImI}{S4\xspace}
\newcommand{\figSmassPPyI}{S5\xspace}
\newcommand{\figSfrequency}{S6\xspace}

\begin{document}
\title{Ion pairing controls rheological properties of ``processionary'' polyelectrolyte hydrogels}
\author{Hassan Srour}
\author{Martien Duvall Deffo Ayagou}
\author{Thi Thanh-Tam Nguyen}
\affiliation{Univ Lyon, Ens de Lyon, Universit\'e Claude Bernard Lyon 1, CNRS,
Laboratoire de Chimie, F-69342 Lyon, France.}
\author{Nicolas Taberlet}
\author{S\'{e}bastien Manneville}
\affiliation{Univ Lyon, Ens de Lyon, Universit\'e Claude Bernard Lyon 1, CNRS,
Laboratoire de Physique, F-69342 Lyon, France.}

\author{Chantal Andraud}
\author{Cyrille Monnereau}
\email{cyrille.monnereau@ens-lyon.fr}
\affiliation{Univ Lyon, Ens de Lyon, Universit\'e Claude Bernard Lyon 1, CNRS,
Laboratoire de Chimie, F-69342 Lyon, France.}

\author{Mathieu Leocmach}
\email{mathieu.leocmach@univ-lyon1.fr.}
\altaffiliation[Twitter: ]{@LamSonLeoc}
\affiliation{Univ Lyon, Universit\'e Claude Bernard Lyon 1, CNRS, Institut Lumi\`ere Mati\`ere, F-69622, VILLEURBANNE, France.}
\begin{abstract}
We demonstrated recently that polyelectrolytes with cationic moieties along the chain and a single anionic head are able to form physical hydrogels due to the reversible nature of the head-to-body ionic bond. Here we generate a variety of such polyelectrolytes with various cationic moieties and counterion combinations starting from a common polymeric platform. We show that the rheological properties (shear modulus, critical strain) of the final hydrogels can be modulated over three orders of magnitude depending on the cation/anion pair. Our data fit remarkably well within a scaling model involving a supramolecular head-to-tail single file between cross-links, akin to the behaviour of pine-processionary caterpillar. This model allows the quantitative measure of the amount of counterion condensation from standard rheology procedure.
\end{abstract} 

\maketitle

\begin{figure*}
\includegraphics{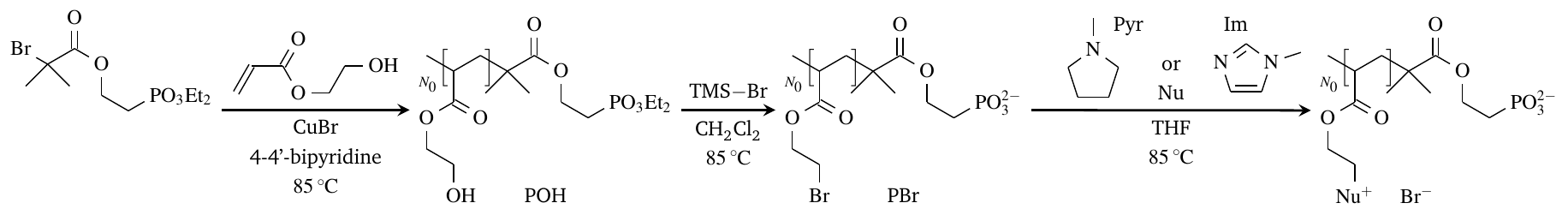}
\caption{Synthesis of \ce{PIm+Br-} and \ce{PPyr+Br-} and their intermediates \ce{POH} and \ce{PBr} with $N_0=70$.}
\label{sch:synthesis}
\end{figure*}
\section{Introduction}
Since the pioneering work of \citet{Wichterle1960} which established their relevance in a biomedical context, polymer-based hydrogels have never ceased to be a very active field of research\cite{Calo2015,Hoffman2001}. They have recently experienced a burst of interest among the biomedical community as controlled drug delivery cargoes or scaffolds for wound healing or tissue repair\cite{Vashist2014,Ratner2004}. Polyelectrolytes are being extensively put to use in this particular context\cite{Rosso2003,BinImran2014}. Their gel formation mechanism often involves reversible-by-nature electrostatic interactions, which can be used for instance to trigger ``smart'' release of bioactive substances\cite{Khare1993,Lockwood2007}. Besides, poly(cationic) gels have been reported to combine antimicrobial activity with scaffold properties for the adherence and growth of cells, and are therefore increasingly used in clinical applications\cite{Hoffman2001,Landers2002}. In this framework, injectability is a key feature, as it provides an easier way to gel delivery in vivo\cite{Tibbitt2016}; thus, reversible shear thinning biocompatible hydrogels are a particularly sought after class of materials\cite{Yu2008}.

Contrary to a small electrolyte for which full ion pairs dissociation is generally achieved in water, polyelectrolytes in solution are in general not fully dissociated, with a substantial fraction of the counterions bound to the polymer\cite{Manning1979}. The resulting net charge of the chain governs the physical properties of the polyelectrolytes solution~\cite{DeGennes1976,Khokhlov1980,Raphael1990,Dobrynin1995,Rubinstein1996}, first of all the ability of the polymer to dissolve in a poor solvent. Counterion condensation is the physical bounding or adsorption of counterions near the polymer chain. Factors influencing this process have been known since the 1880's but are still not fully understood. \citet{Hofmeister1888} was the first to propose a systematic ranking of ions, based on their propensity to promote the coalescence of egg white. This so-called Hofmeister series has since proved quite universally valid\cite{Zhang2010a}, including multi-charged polymers such as naturally occurring proteins or synthetic polyelectrolytes as well as charged colloidal particles\cite{Schwierz2010,Oncsik2015} or soft matter interfaces in general\cite{Leontidis2014}. However, depending on the nature of the polymer (hydrophilic or hydrophobic, anionic or cationic), the Hofmeister series can be direct (in short, well hydrated ions promote condensation) or reverse~\cite{Schwierz2010,Oncsik2015}.

We have recently reported on a highly reversible poly(cation) based hydrogel\cite{Srour2014}, which formation relies on a new concept of supramolecular  electrostatic interaction. Briefly, the atom transfer radical polymerisation (ATRP) is initiated by a phosphonate-terminated ATRP initiator, eventually affording an \emph{anion-terminated poly(cationic) polymer}. When dispersed in an aqueous medium, this polymer forms an hydrogel through the occurrence of a head(anion)-to-body(cations) supramolecular network. This highly dynamic electrostatic network provides the resulting gel with spectacular mechanical and self-healing properties. Moreover the interaction can be disrupted by various chemical stimuli, such as pH or ionic strength.

In the present report, we investigate the role of the counterion condensation on the mechanical properties of such hydrogels. In a first part we describe how we take advantage of our postfunctionalization approach to play systematically with the nature (aromatic or not) of the cationic repeating unit and the associated halide counterion (\ce{F-}/\ce{Cl-}/\ce{Br-}/\ce{I-}) varied along the Hofmeister series. In the second and third parts, we show respectively the qualitative change in gel formability and the quantitative variations of mechanical properties of the aqueous dispersions obtained from these well-characterized polymers. In a fourth part, we rationalize this behaviour by proposing a microscopic model based on the idea that at low dissociation rates, cross-links (defined as a point where three or more polymers meet) are not separated by a single macromolecule but by several (up to hundreds) polymers in a supramolecular chain. We call this behaviour ``processionary'' in analogy to pine processionary caterpillar (\textit{Thaumetopoea pityocampa}) behaviour\cite{Fabre1916}. In particular, this model enables the quantification of the charge condensation rate from standard rheological measurements.

\section{Results and discussion}

\subsection{Synthesis and characterizations}

We synthesise the phosphonate terminated polymer hereafter referred to as \ce{PBr} according to our previously reported methodology\cite{Srour2014,Appukuttan2012}. We obtain a well-controlled linear polymer ($M_w= \SI{8200}{\dalton}$ from NMR, \SI{5614}{\dalton} from GPC with $M_w/M_n = 1.08$; see Supplementary Materials and Supplementary Figure~\figSsec), with a degree of polymerisation $N_0=70$ (NMR). This polymer serves as a common scaffold from which we derive  all studied systems. Nucleophilic addition of N-methylimidazole or N-methylpyrrolidine to a heated solution of \ce{PBr} in THF affords the corresponding poly(imidazolium) and poly(pyrrolidinium) compounds. In these cases, as an inherent consequence of the structure of the starting material, charge neutrality is provided by bromide counter ions, and the polymer will be referred to as \ce{PIm+Br-} and \ce{PPyr+Br-}, respectively (Figure~\ref{sch:synthesis}).

\begin{figure}
\includegraphics{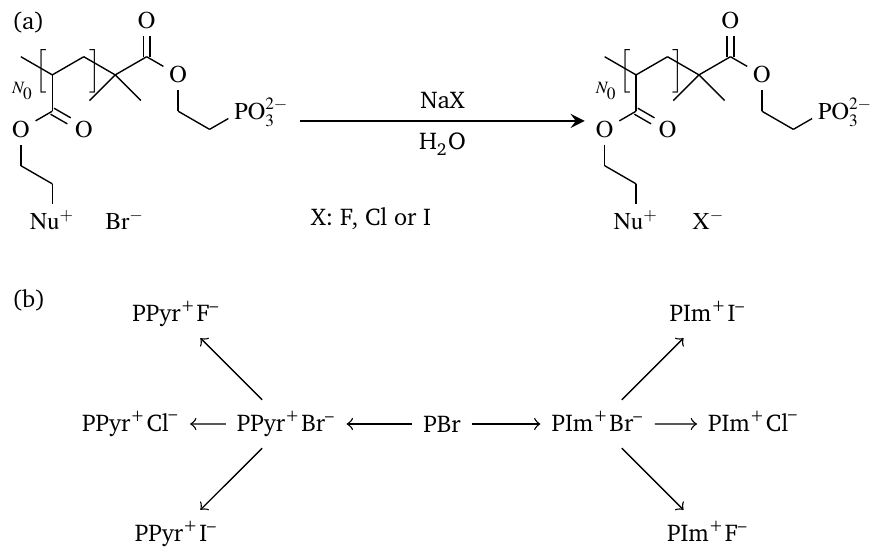}
\caption{(a) Anionic metathesis reaction. As above, \ce{Nu+} corresponds either to pyrrolidinium or imidazolium. (b) Summary of the polymers derived from a single batch of \ce{PBr}.}
\label{sch:metathesis}
\end{figure}

We performed anionic metathesis by pouring an aqueous solution of \ce{PIm+Br-} or \ce{PPyr+Br-} into a saturated aqueous solution of the different sodium halides (\ce{NaF}, \ce{NaCl}, \ce{NaI}) (Figure~\ref{sch:metathesis}). In the cases of \ce{PIm+F-}, \ce{PIm+Cl-} , \ce{PPyr+F-} and \ce{PPyr+Cl-}, we obtain a turbid suspension immediately after addition. After extensive dialysis of the resulting mixture against deionized water and lyophilization we recover in high yields the different \ce{PIm+X-} and \ce{PPyr+X-} (where \ce{X}=\ce{F}, \ce{Cl}, \ce{I}). In order to bring evidence for the efficiency of the ionic metathesis, we submit the resulting materials to negative-mode high-resolution mass spectrometry (HRMS, see Supplementary Materials). We unambiguously assess complete displacement of bromide counterions by the full disappearance of the diisotopic mass peak (89/\SI{91}{\dalton}, Supplementary Figures~\figSmassPImBr and \figSmassPPyBr). We further confirm substitution by iodide by the concomitant apparition of a characteristic monoisotopic peak (\SI{121}{\dalton}, Supplementary Figures~\figSmassPImI and \figSmassPPyI), while fluoride and chloride anions signals, which are below the detection limits of the spectrometer, could not be observed in the corresponding polymers (not shown).

\begin{figure}
\includegraphics{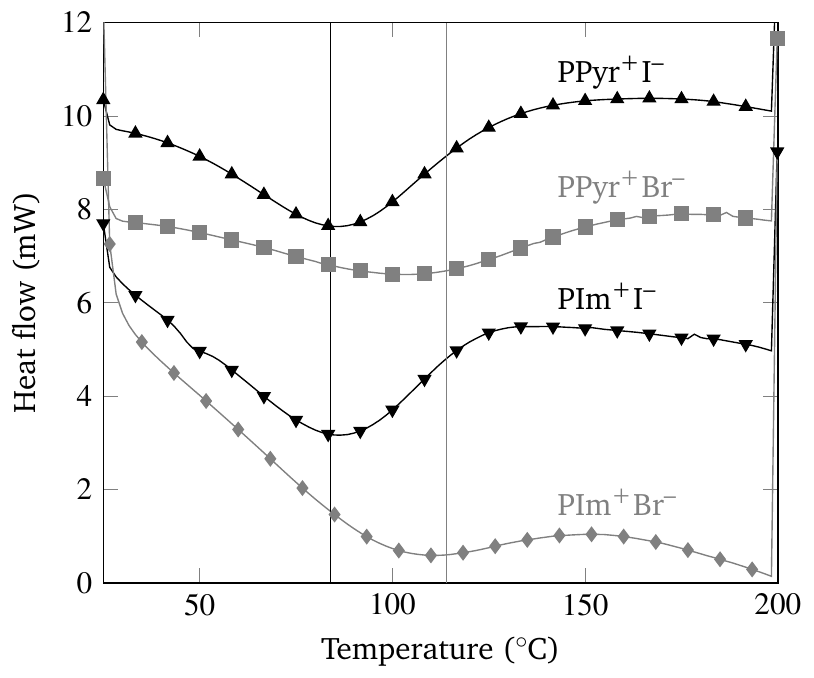}
\caption{DSC measurements of polymer samples showing the presence of an endothermic peak. Curves are shifted vertically by arbitrary amounts for clarity. Vertical lines indicate the peak temperatures at \SI{84}{\celsius} for the two iodides in black and \SI{114}{\celsius} for the two bromides in gray.}
\label{fig:dsc}
\end{figure}

Differential Scanning Calorimetry (DSC) thermograms  of all synthesized polymers also reveals marked differences in their properties, that can only be explained by differences in their counterion features. Thus, in the case polyelectrolytes with fluoride and chloride counterion, we observe no exo or endothermic transition below \SI{200}{\celsius} (featureless thermograms, not shown). By contrast, we observe a broad endothermic peak when iodide and bromide are used as counterions, see Figure~\ref{fig:dsc}. Quite remarkably, we found a similar peak temperature (84-\SI{85}{\celsius}) for \ce{PIm+I-} and \ce{PPyr+I-}. When bromide counterions are present, we observe a significant increase of the peak temperature (114-\SI{115}{\celsius}) but again, with a similar value between \ce{PIm+Br-} and \ce{PPyr+Br-} as shown in Figure~\ref{fig:dsc}. The overall shape and position of the DSC peaks is very reminiscent of previously reported data for various naturally occurring or synthetic poly(electrolytes)\cite{Li2005,Sarmento2006,Ostrowska-Czubenko2009a,Moin2015}. It is generally attributed to desorption of weakly bound water from the polymer network. 

This difference in DSC profile can be understood by considering the water-binding ability of the different polymers in this study. Since the peak temperature does not depend on the presence of imidazolium or pyrrolidinium, water-binding is probably due to the only part of the side groups able to accept hydrogen bonds from water: the ester function. Following \citet{Zhang2010a}, we propose that these bonds can be enhanced if the involved water molecule is polarized by solvating an anion. The polarization, and thus the water-binding ability will decrease along the direct Hofmeister series $\ce{F-}>\ce{Cl-}>\ce{Br-}>\ce{I-}$. We are able to observe the end of this trend in the shift of peak position between bromide and iodide counterions, whereas in the case of chloride and fluoride, binding is too strong to observe water desorption. This hypothesis is further confirmed below, when looking at the properties of different polymers in aqueous dispersion.

\subsection{Water swelling, counterion condensation and solvent quality}
Water-swelling properties turn out to be extremely dependent on the polymer composition. With bromide or iodide conterions, both poly(pyrrolidinium) and poly(immidazolium) polymers afford homogeneous and optically transparent gels upon swelling with deionised water. We observe a limited swelling for \ce{PPyr+Cl-}, resulting in a granular, inhomogeneous gel. With fluoride counterions, or in the case of \ce{PIm+Cl-}, we observe neither swelling nor dissolution. We recover a 7.4~\%~wt suspension as a biphasic mixture of its individual components.

The above dependence of solubility on the nature of the counterions follows the direct Hofmeister series. The strong condensation of \ce{F-} on the chain decreases the effective positive charge of the polymer and thus reduces solubility. Conversely, bulkier and softer ions like \ce{I-} are less adsorbed, resulting in more dissociated charges and higher solubility. As a consequence, while dissolution is observed for the four polymers with bromide and iodide counterions, in which ion dissociation takes place to a sufficient extent, the two polymers with fluoride counterions, for which counterion condensation is expected to be strong, cannot be dissolved.

However Hofmeister series alone cannot explain the difference between soluble \ce{PPyr+Cl-} and insoluble \ce{PIm+Cl-}. This constitutes a strong indication that counterions are more strongly condensed near imidazolium than near pyrrolidinium. Indeed, although both cationic in nature, pyrrolidinium and imidazolium ions have markedly different properties. Because of their aromatic ring, imidazolium ions are particularly prone to promote a variety of supramolecular interactions which strongly contribute to their physicochemical properties, such as \ce{\pi+-\pi} or anion$-\pi^+$.  More recently, \ce{\pi+-\pi+} interactions have also been identified as a distinctive driving force for imidazolium dimerization\cite{Geronimo2011}. Because of strong intercoulombic repulsion, \ce{\pi+-\pi+} interactions are primarily weaker than their \ce{\pi-\pi} counterparts. However it has been established that, in the presence of counterions, the stabilizing effect of \ce{\pi+-\pi+} interaction could reach magnitudes largely exceeding that of the latter\cite{Geronimo2011}. In particular, dimerization of imidazolium-chloride ion pairs have been the object of recent studies, and it has been shown that the involvement of the negatively charged counterion was essential in maximizing the stabilization of the \ce{\pi+-\pi+} complex by minimizing coulombic repulsion between the imidazolium moieties\cite{Matthews2014,Gao2015a}. It is likewise well documented that covalent incorporation of interacting groups within a polymer chain strongly favours intramolecular interactions between these groups, by increasing their so-called effective molarity\cite{Li2003,Mulder2004,Huerta2013}. It is therefore not surprising that both effects participate in making the imidazolium-anion interaction stronger than pyrrolidinium at the detriment of solubility.

Such an insolubility in absence of charge dissociation indicates that the chain is in poor solvent. Locally, the monomers are condensed in collapsed globules to minimize the contacts with water molecules, and solubility on larger scales can only be achieved if the polymer bears enough charges\cite{Khokhlov1980,Raphael1990}. To estimate the $\Theta$-temperature, that delimit poor and good solvent behaviour, we held the two fluoride polymers in water at boiling temperature overnight. No dissolution or swelling was observed, implying that $\Theta>\SI{100}{\celsius}$.

\subsection{Rheological measurements}
\begin{figure*}
\includegraphics{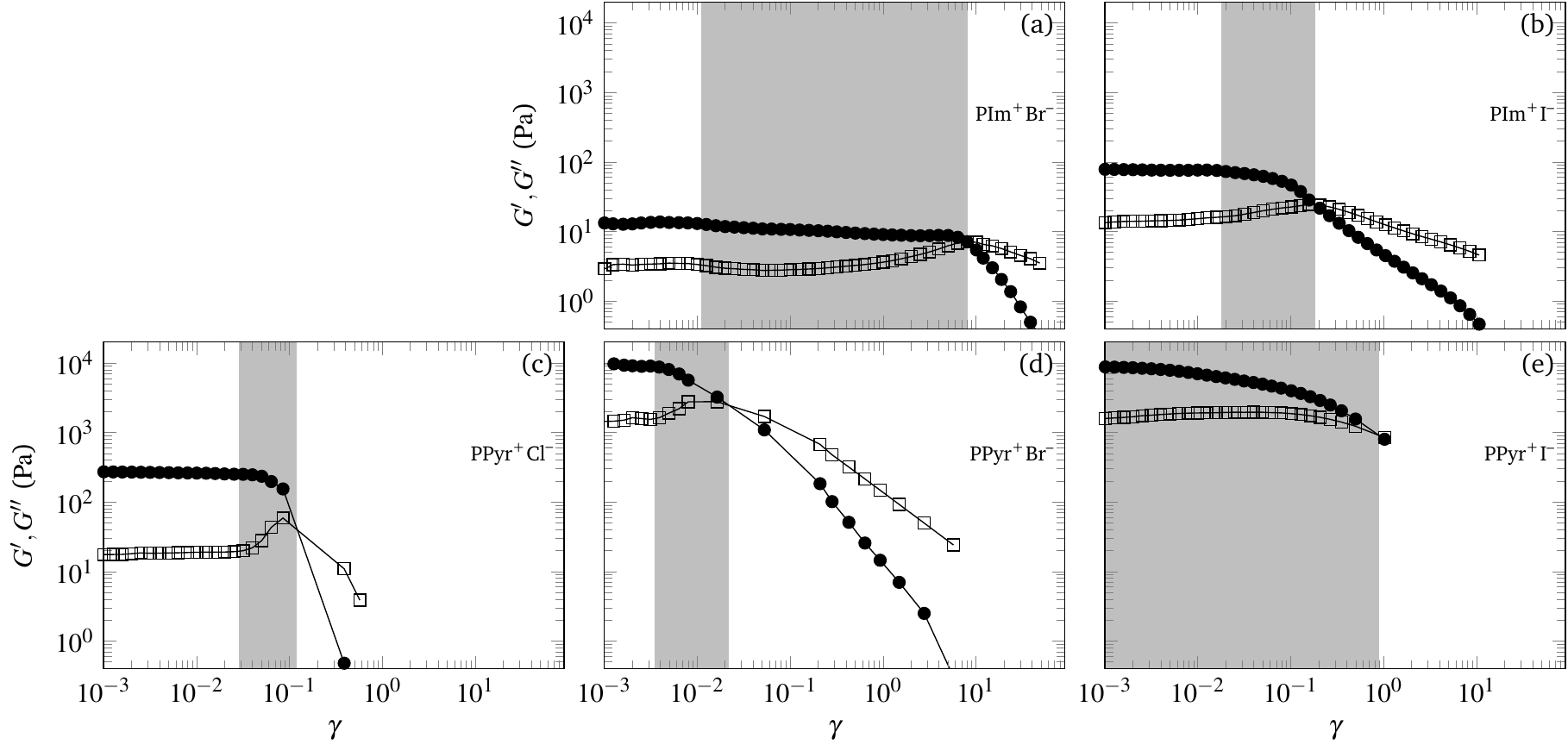}%
\caption{Storage modulus $G^\prime$ ($\bullet$) and loss modulus $G^{\prime\prime}$ ($\square$) measured through oscillatory shear experiments plotted against the strain amplitude $\gamma$. The moieties change with rows and the counterions with columns. The fixed frequency is $f=\SI{1}{\hertz}$. All samples are at 7.4~\%~wt. The gray area encompasses the plastic regime that lies between the linear regime at very low amplitudes and the fluid-like shear-thinning regime at high strain amplitude.}%
\label{fig:strain}%
\end{figure*}

We perform rheological studies of the water-swelling materials. Briefly, data are recorded at \SI{25}{\celsius} with an AR 1000 rheometer (TA Instruments) in a cone-plate geometry of radius \SI{40}{\milli\metre}, an angle of \SI{2}{\degree} and a truncation of \SI{58}{\micro\metre}\cite{Macosko1994,Larson1999}. We place the sample on the plate, and then lower the cone to the measuring position, spreading the sample in the process. We remove excess material, so that the sample exactly fills the gap. Due to Weissenberg effect at high shear rate, we apply no preshear before measurement. To minimize water absorption, we cover the geometry with a solvent trap, using light mineral oil as a liquid seal between the rotor and the cap. After one minute of equilibration we perform an oscillatory frequency sweep at small amplitude (strain amplitude $\gamma=0.1\%$, see Supplementary Figure~\figSfrequency) and then an oscillatory strain sweep at fixed frequency ($f=\SI{1}{\hertz}$). For fluoride counterions as well as \ce{PIm+Cl-}, the rheological profile is that of pure water which confirms the visual observation. However all other samples behave as shear-thinning yield stress fluids, confirming our previous results with \ce{PIm+Br-}~\cite{Srour2014}. As shown in Figure~\ref{fig:strain}, soluble samples are solid-like ($G^\prime \gg G^{\prime\prime}$) at small strain and flow at large strains ($G^{\prime\prime} \gg G^\prime$) with a steep decrease of the moduli. We confirm solid-like behavior at low strain for all accessible frequencies (see Supplementary Figure~\figSfrequency). We checked that the linear mechanical properties are unchanged by the flow history given a few minutes of rest.

The mechanical properties at low strain evolve with the factors influencing counter-ion condensation. Imidazolium based gels are much weaker than their pyrrolidinium counterpart. Comparison between \ce{PIm+Br-} and \ce{PPyr+Br-} is particularly illustrative of this trend, as their $G^\prime$ values (\SI{13}{\pascal} and \SI{8100}{\pascal}, resp.) differ by almost three orders of magnitude. On the bottom row of Figure~\ref{fig:strain} we can compare the rheological responses of poly(pyrrolidinium)-based hydrogels for the three counterions allowing dissolution in water. Gels with bromide and iodide counterions show roughly the same modulus value at vanishing strain for both $G^\prime$ and $G^{\prime\prime}$. By contrast, the gel with chloride counterions is weaker by more than one order of magnitude. In other words, counterion condensation is correlated to softer gel. This trend is confirmed between \ce{PIm+Br-} and \ce{PIm+I-}, the former being an order of magnitude softer than the later.

Moderately charged \ce{PIm+I-} and \ce{PPyr+Cl-} show a plateau in both moduli at low strain corresponding to the linear regime of the material; an overshoot of $G^{\prime\prime}$ and a downward slope of $G^\prime$ at intermediate strains corresponding to the onset of plasticity~\cite{Hyun2011}; and a decrease of both moduli at large strains, steeper for $G^\prime$ than for $G^{\prime\prime}$, indicating shear thinning. The linear regime is either extremely narrow or non existent in heavily charged \ce{PPyr+Br-} and \ce{PPyr+I-}. By contrast the lightly charged \ce{PIm+Br-} displays a broad plastic regime (tree decades of strain) between the linear regime and the crossing of the moduli. In any case, the end of the linear regime corresponds to a strain much smaller than 1.

In the following we show that the correlation between softness and counterion condensation can be explained on the basis of a microscopic model of interchain interactions.

\subsection{Processionary model}

\begin{figure}
\centering\includegraphics{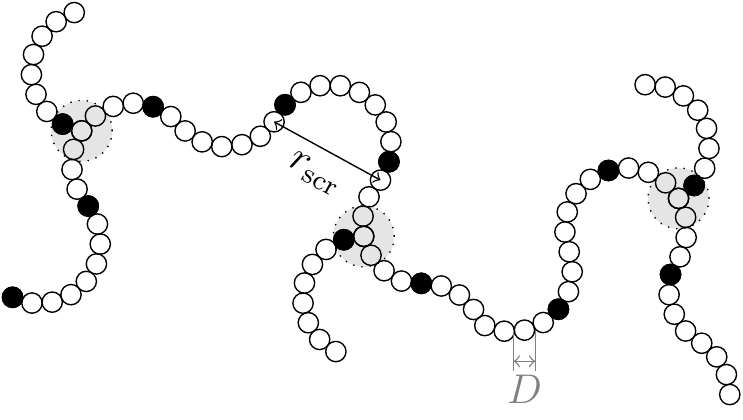}
\caption{Processionary model. Sketch of the network in the case of a procession of size $n=3$ between cross-links (gray disks) and a persistence length $r_\mathrm{scr}\approx 6D$. Empty circles are not individual monomers but electrostatic blobs of size $D$. Electrostatic blobs containing an anionic head are shown as black filled circles.}
\label{fig:procession}
\end{figure}
\begin{figure*}
\centering\includegraphics{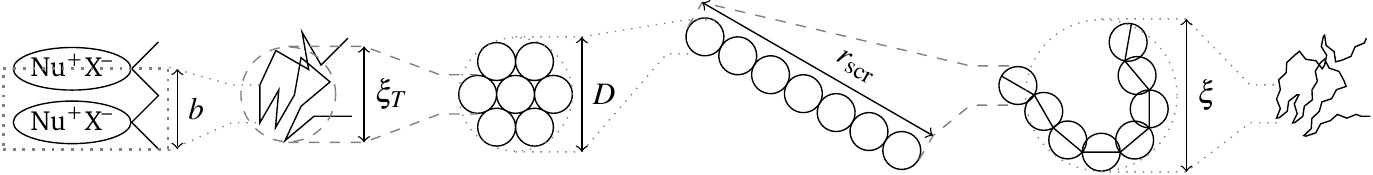}
\caption{Conformation of semi-dilute polyelectrolytes at rest at various scales. Length increases from left to right.}
\label{fig:scales}
\end{figure*}

We have previously demonstrated that, due to the opposite charges of the phosphonate head, head-to-body ionic bonds are possible~\cite{Srour2014}. On the one hand, if at least two foreign heads attach to the same body, we obtain an effective cross-link point. When the probability of such a configuration is non-zero, we obtain a physically cross-linked gel. This is the ideal situation that we described in our previous paper.

On the other hand, if every chain has a single foreign head attached to its body, every polymer is linked to two others in a single file\footnote[3]{We neglect the possibility of the two charges of the phosphonate head to attach to two separate chains}. In addition, if we suppose that in order to minimize the inter-chain repulsion between charged cationic groups, heads are preferentially attached to the tail of their neighbour, we obtain a linear chain of effective polymerisation index $n N_0$ where $n$ is the number of polymers in the supramolecular queue. This behaviour is somewhat evocative of that of the pine processionary caterpillar memorably described by \citet{Fabre1916}: 
``They proceed in single file, in a continuous row, each touching with its head the rear of the one in front of it. [...] No Greek theoria winding its way to the Eleusinian festivals was ever more orderly. Hence the name of Processionary given to the gnawer of the pine.''

We thus have two limiting cases: (i) every chain has at least two heads attached and we have roughly $N_0$ monomers between cross-link points; (ii) every chain has at the most a single head attached and we have isolated supramolecular chains in the system. The former case should be observed when a significant number of ion-pairs along the polymer chain are dissociated providing a significant probability for multiple phosphonate heads binding on a single polymer chain. Conversely, the latter case should be observed at low charge dissociation. In between these two limiting cases, we should observe cross-link points separated by processions of $n N_0$ monomers as sketched in Figure~\ref{fig:procession}.

As we will discuss in the following, all rheological features of the hydrogels studied here can  be rationalised on the basis of this ``processionary'' model. We base our analysis on three main observables, see Table~\ref{tab:results}: (i) the value of the shear modulus at small amplitude $G^\prime(\gamma\rightarrow 0)$, i.e. the elasticity of the undamaged gel network, (ii) the strain amplitude $\gamma_0$ corresponding to the end of the linear regime~\cite{Hyun2011}, and (iii) the strain amplitude $\gamma_c$ corresponding to the crossing of the moduli. By using these three parameters, we show that it is possible, with reasonable assumptions, to estimate microscopic parameters such as the average number of chains between cross-links $n$, the number of monomers between dissociated ion pairs $A$ that quantify counterion condensation and the head-to-body bonding energy $E_c$.

\subsubsection{Conformation at rest.}
The scaling theory of \citet{Dobrynin1995} describes the structure of a polyelectrolyte chain. In the following, we make the link between this theory and the chemical structure of our polymers. In absence of strain, the structure of a polyelectrolyte chain is organised at various scales, sketched on Figure~\ref{fig:scales}.

\paragraph{Kuhn length.}
The smallest scale is the Kuhn length $b$ containing $g_\mathrm{K}$ monomers. For a freely joint chain of tetrahedral carbons $b\approx 367$ pm which corresponds to 3 tetrahedral carbons. Since a monomer counts 2 carbons along the chain $g_\mathrm{K}\approx 3/2$.

\paragraph{Thermal length.}
At the thermal length $\xi_T$, the attractive potential between monomers is balanced by thermal energy $k_\mathrm{B}T$, with $k_\mathrm{B}$ the Boltzmann constant. This balance of energy can be written function of the reduced temperature $\tau = 1 -T/\Theta$ and the number $g_T$ of monomers in the thermal blob:
\begin{equation}
k_\mathrm{B}T = \left(\frac{b}{\xi_T}\right)^3\left(\frac{g_T}{g_K}\right)^2 \tau k_\mathrm{B}T 
\label{eq:poorkT}
\end{equation}
Between $b$ and $\xi_T$ we have a persistent random walk $\xi_T = b (g_T/g_\mathrm{K})^{1/2}$. These conditions yields 
\begin{equation}
\xi_T = b/\tau,\qquad g_T = g_\mathrm{K}/\tau^2
\label{eq:thermal}
\end{equation}

\paragraph{Electrostatic length, weak charging case.}
Following \citet{Khokhlov1980} one can further define a third characteristic length $D$ over which the electrostatic energy become dominant over short range attraction or surface energy of the collapsed polymer. $D$ defines the size of the electrostatic blob and we name $g_e$ the number of monomers in it. Due to counterion condensation, not all monomers are charged. Counterion condensation is quantified by assuming that there is a charge every $A$ monomers. Therefore there are $g_e/A$ charges in the electrostatic blob and the electrostatic energy reads $E_e = (g_e/A)^2 e^2/(4\pi\epsilon D)$, with $e$ the elementary charge and $\epsilon$ the dielectric constant of the solvent. If $D>\xi_T$ the main opposition to the electrostatic energy can be understood as a surface energy proportional to the number of thermal blobs exposed on the surface $E_s = k_\mathrm{B}T (D/\xi_T)^2$. The balance of energies yields:
\begin{equation}\left(\frac{D}{\xi_T}\right)^2 = \left(\frac{g_e}{A}\right)^2 \frac{\ell_\mathrm{B}}{D},
\label{eq:electrosurface}
\end{equation}
where $\ell_\mathrm{B} = e^2/(4\pi\epsilon k_\mathrm{B}T)$ is the Bjerrum length. In water $\ell_\mathrm{B} \approx \SI{0.7}{\nano\metre}$. Since thermal blobs fill the volume of the electrostatic blob, one has $D = \xi_T \left(g_e/g_T\right)^{1/3}$ and using Equation~\eqref{eq:thermal} we obtain
\begin{equation}
g_e = \frac{A^2}{u g_\mathrm{K}} \tau, \qquad D = b\left(\frac{A^2}{u g_\mathrm{K}^2}\right)^{1/3},\qquad \text{with }u = \ell_\mathrm{B}/b.
\label{eq:geD}
\end{equation}

\citet{Dobrynin1995} introduced the extension parameter $B$ defined as the ratio between the length of a fully extended chain of $g_e$ monomers ($g_e/g_\mathrm{K}$ Kuhn segments) and the actual size of the electrostatic blob: 
\begin{equation}
B = \frac{g_e}{g_\mathrm{K}}\frac{b}{D} = \left(\frac{A^2}{u g_\mathrm{K}^2}\right)^{2/3} \tau
\label{eq:B}
\end{equation}

At constant solvent quality and polymer architecture, $B$ is thus monotonically related to the amount of counterion condensation $A$. In the following we will use $B$ as our main variable and estimate it from rheological measurements. We conveniently combine Equations~(\ref{eq:geD}) and (\ref{eq:B}) to express $g_e$ and $D$ function of $B$:
\begin{equation}
g_e = \left(\frac{B^3}{\tau}\right)^{1/2} g_\mathrm{K},\qquad D = b \left(\frac{B}{\tau}\right)^{1/2}.
\label{eq:geDfromlargeB}
\end{equation}

\paragraph{Electrostatic length, strong charging case.}
From Equations~(\ref{eq:thermal}) and (\ref{eq:geDfromlargeB}) we observe that for $B<1/\tau$ the size of the electrostatic blob should be smaller than the thermal length and the assumptions behind Equation~(\ref{eq:electrosurface}) break down. To our knowledge the study of polyelectrolytes in poor solvent by \citet{Khokhlov1980} and subsequent literature~\cite{Dobrynin1995,Rubinstein1996} focused on the weak charging regime and did not treat the small $B$ regime. When $B\ll 1/\tau$, most of the counterions are not condensed and attraction potential between monomers competes directly against the electrostatic repulsion. The size of the electrostatic blob is given by the following balance of energies:
\begin{equation}
\left(\frac{b}{D}\right)^3\left(\frac{g_e}{g_K}\right)^2 \tau = \left(\frac{g_e}{A}\right)^2 \frac{\ell_\mathrm{B}}{D}
\label{eq:poorelec}
\end{equation}
yielding, with $D = b(g_e/g_\mathrm{K})^{1/2}$,
\begin{equation}
D = b \left(\frac{A^2\tau}{u g_\mathrm{K}^2}\right)^{1/2}, \qquad g_e = \frac{A^2\tau}{u g_\mathrm{K}},
\end{equation}

In this weak counterion condensation regime, the definition of the extension parameter $B$ yields a different relation with $A$ and both $D$ and $g_e$:
\begin{equation}
B = \frac{g_e}{g_\mathrm{K}}\frac{b}{D} = \left(\frac{A^2\tau}{u g_\mathrm{K}^2}\right)^{1/2}, \qquad D = bB, \qquad g_e = B^2 g_\mathrm{K}
\label{eq:geDfromsmallB}
\end{equation}

Here we shall describe the crossover between weak and strong counterion condensation regimes by combining Equations~\eqref{eq:geDfromlargeB} and \eqref{eq:geDfromsmallB} into
\begin{equation}
D = bB (1+B\tau)^{-1/2}, \qquad
g_e = g_\mathrm{K} B^2 (1+B\tau)^{-1/2},
\label{eq:geDfromB}
\end{equation}
and we generalize the relation between $A$ and $B$ as
\begin{equation}
A = u^{1/2} g_\mathrm{K} \frac{B}{\tau^{1/2}}(1+B\tau)^{-1/4}.
\label{eq:A}
\end{equation}

\paragraph{Screening length.}
The next length scale is the screening length $r_\mathrm{scr}$. Between $D$ and $r_\mathrm{scr}$ the electrostatic blobs are organised in a linear rod containing $g_\mathrm{scr}$ monomers. A rod is $B$ times shorter than the fully extended $g_\mathrm{scr}/g_\mathrm{K}$ Kuhn segments that it contains such that
\begin{equation}
r_\mathrm{scr}= g_\mathrm{scr} b / (B g_\mathrm{K}).
\label{eq:rod}
\end{equation}
To obtain $r_\mathrm{scr}$ we follow \cite{Dobrynin1995} by considering in a first step the case of a dilute solution where the length $L$ of a single chain of $\mathcal{N}$ monomers is shorter than the screening length. Equation~(\ref{eq:rod}) becomes $L= \mathcal{N} b / (B g_\mathrm{K})$. Therefore the overlap concentration is $c^* = \mathcal{N}/L^3 = B g_\mathrm{K} / (b L^2)$. In a second step, when the monomer concentration $c$ is larger than $c^*$ and in absence of salt, the screening length is
\begin{equation}
r_\mathrm{scr} = L \left(\frac{c^*}{c}\right)^{1/2} =  \left(\frac{B g_\mathrm{K}}{cb}\right)^{1/2}.
\label{eq:rscrNoSalt}
\end{equation}

When charge screening is mainly due to the presence of added salt at a concentration $c_s$ much larger than the dissociated counterions, electrostatic interactions are screened at the Debye length $r_\mathrm{scr} = \left(\ell_\mathrm{B} c_s\right)^{-1/2} \equiv \kappa^{-1}$. Finally, for arbitrary salt concentration \citet{Dobrynin1995} use the crossover expression,
\begin{equation}
r_\mathrm{scr} = \left(\frac{B g_\mathrm{K}}{cb}\right)^{1/2} \left(1+ B u \frac{c_s}{c} g_\mathrm{K}\right)^{-1/2}.
\label{eq:rscr}
\end{equation}

From Equations~(\ref{eq:rod}) and (\ref{eq:rscr}) we obtain
\begin{equation}
g_\mathrm{scr} = \left(\frac{B^3 g_\mathrm{K}^3}{cb^3}\right)^{1/2} \left(1 + B u \frac{c_s}{c} g_\mathrm{K}\right)^{-1/2}.
\label{eq:gscr}
\end{equation}

\paragraph{Correlation length.}
The last length scale is the correlation length $\xi$. Between $r_\mathrm{scr}$ and $\xi$ the chain forms a self avoiding walk of persistence length $r_\mathrm{scr}$. Above the correlation length the polyelectrolyte chain forms a random walk of correlation blobs containing $g=c\xi^3$ monomers.

\subsubsection{From modulus to procession length.}

Let us note $N$ the number of monomers between two crosslinks or entanglements. If $N>g$ the ``procession'' of polymer chains performs a random walk of correlation blobs, each being a self-avoiding walk. If $g>N>g_\mathrm{scr}$ the procession performs only a self-avoiding walk. In any case, each procession is thus an entropic spring of constant stiffness $k_\mathrm{B}T$. The number density of procession is $c/N$. Therefore, the shear modulus at low strain is given by:
\begin{equation}
G = \frac{c}{N}k_\mathrm{B}T.
\label{eq:G}
\end{equation}
Because of the different weights of the pyrrolidinium and imidazolium moieties and of the different counterions, $c$ is not constant as we chose to conduct our experiments at constant weight fraction $w$ of polymer (generally used to quantify gelation ability of a given gelator). This explains why the respective modulus or critical strains of \ce{PPyr+I-} (heavier, lower number density) and \ce{PPyr+Br-} are in reverse order with respect to the Hofmeister series. Taking into account the molecular mass $M$ of each polymer, the Avogadro number $\mathcal{N_A}$ and  knowing the density $d$ of the solvent, we obtain the number $n = N/N_0$ of chains between cross-link
\begin{equation}
n = \frac{\mathcal{N_A}}{M} w d \frac{k_\mathrm{B}T}{G}.
\label{eq:n}
\end{equation}

We find that the number $n$ of chains between cross-link point goes from 1 in \ce{PPyr+I-} and \ce{PPy+Br-} to 800 in \ce{PIm+Br-}, following the \textit{a priori} ranking of charge dissociation, see Table~\ref{tab:results}. This larger cross-link ratio is consistent with a higher probability of attaching two or more heads on a highly dissociated body.

\begin{table*}
\sisetup{
separate-uncertainty,
}\centering
\begin{tabular}{@{}lSSSSSSSSS@{}}\hline
& {$G^\prime(\gamma\rightarrow 0)$} & {$\gamma_0$} & {$\gamma_c$} & {$n$} & {$B$} & {$E_c$} & {$E_c$} & {$D$} & {$A$}\\
&	{\si{\pascal}} & {\%} & {\%} & & & {\si{\kilo\joule\per\mole}}& {$k_\mathrm{B}T$} & {$\ell_B$}&\\\hline&&&&&\\[-10pt]
\ce{PIm+Br-}	&	13	&	1.1	&	800	&	800	&	830	&	304	&	123	&	24	&	640\\
\ce{PIm+I-}	&	78	&	1.8	&	18	&	113	&	113	&	116	&	46.8	&	8.7	&	140\\
\ce{PPyr+Cl-}	&	272	&	2.8	&	10	&	46.8	&	58.2	&	66	&	26.6	&	6.2	&	90\\
\ce{PPyr+Br-}	&	9810	&	0.3	&	2	&	1.1	&	2.6	&	5.2	&	2.1	&	1.0	&	7.1\\
\ce{PPyr+I-}	&	8800	&	0.1	&	90	&	1.0	&	2.4	&	5.0	&	2.0	&	0.9	&	6.5\\
\hline
\end{tabular}
\caption{Summary of rheological measurements and microscopic values deduced from the model. $E_c$, $D$ and $A$ are obtained by assuming $\tau=0.40$}
\label{tab:results}
\end{table*}

\subsubsection{Limit of the linear regime.}

\begin{figure}
\centering\includegraphics{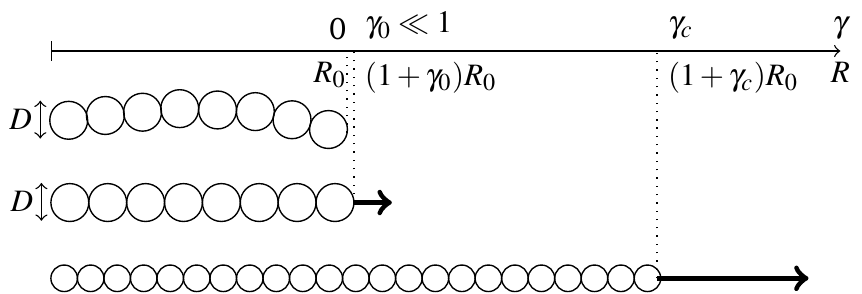}
\caption{Shearing the gel translates into stretching individual processions. From top to bottom: configuration at rest; full extension of the self avoiding walk; stretching of the electrostatic blobs.}
\label{fig:stretch}
\end{figure}

As the material is strained, individual processions are stretched, starting from the larger scales~\cite{Pincus1976}. As sketched in Figure~\ref{fig:stretch}, when the the correlation blobs are fully stretched, the procession is a linear assembly of electrostatic blobs. Stretching the chain further means exposing more thermal blobs to the solvent, a process that cannot be expressed using the model of a spring of constant stiffness. This condition is thus the limit between the linear and non linear regimes.

For all our samples, the linear regime is narrow $\gamma_0\approx 10^{-2}\ll 1$ if existing. It means that the procession is just long enough to start performing the self-avoiding walk. Incidentally, this implies that the procession is too short to perform the random walk. In other words the number $N$ of monomers between crosslinks is comparable to $g_\mathrm{scr}$. Using equation (\ref{eq:gscr}), this condition translates into
\begin{equation}
\left(\frac{B}{B_0}\right)^3 = 1 + \frac{B}{B_s}
\label{eq:solveB}
\end{equation}
with $B_s = c/(u c_s g_\mathrm{K})$ and $B_0 = b c^{1/3} N^{2/3} g_\mathrm{K}^{-1} = (bc/g_\mathrm{K}) \left(k_\mathrm{B}T/G\right)^{2/3}$. Here, we do not have salt per se, however negatively charged phosphonate heads and their counterions play the same role by contributing to screening. We thus have two ``salt'' charges per chain. Thus $c_s/c = 2/N_0$ and $B_s = N_0/(2u g_\mathrm{K}) \approx 12$. $B_0$ is obtained from $G^\prime(\gamma\rightarrow 0)$ for each sample.

Analysing Equation (\ref{eq:solveB}), we identify two physically relevant limit cases: (i) if $B \ll B_s$ screening is mainly due to uncondensed counterions and $B = B_0$; (ii) if $B \gg B_s$ screening is mainly due to phosphonate heads and $B = \left(B_0^3/B_s\right)^{1/2} > B_0$.

Numerical solutions of Equation (\ref{eq:solveB}) are given in Table~\ref{tab:results}. For the same quality of solvent $B$ is monotonically related to the amount of counterion condensation. Since our samples are sorted by increasing $B$ we confirm that they are sorted by decreasing counterion condensation. To be more quantitative, we extract the number $A$ of monomers between dissociated charges from Equation~(\ref{eq:A})
.
We know that $\Theta>\SI{100}{\celsius}$ so $0.2<\tau<1$. In the following we will deduce a more precise measure of $\tau$ and thus values of $A$.

%

\subsubsection{Extension of the electrostatic blobs.}

The fully extended self-avoiding walk is a cylinder of length $R_0 = Nb/(Bg_\mathrm{K})$, of diameter $D$, of volume $V = R_0 D^2$ and of area $\Sigma_0 = R_0 D = (R_0 V)^{1/2}$. Extending the cylinder to $R = (\gamma+1) R_0$ while keeping the volume $V$ constant thus creates an extra area 
$
\Delta\Sigma = \left(\left(\gamma +1\right)^{1/2} -1\right) \Sigma_0
$
that translates into a stretching energy due to the surface tension $k_\mathrm{B}T/\xi_T^2$, via Equations~(\ref{eq:thermal}) and (\ref{eq:geDfromB}):
\begin{equation}
\frac{E_\mathrm{stretch}}{k_\mathrm{B}T} = \left(\left(\gamma +1\right)^{1/2} -1\right) \frac{N}{g_\mathrm{K}} (1+B\tau)^{-1/2} \tau^2.
\label{eq:stretch}
\end{equation}

If we consider that the head-to-body bonds break at $\gamma_c$, the reduced temperature is solution of
\begin{equation}
\left(\left((\gamma_c +1)^{1/2} -1\right)\frac{k_\mathrm{B}T}{E_c}\frac{N}{g_\mathrm{K}}\right)^2 \tau^4 - B\tau -1 = 0
\label{eq:tau}
\end{equation}
where $E_c$ is the energy of the head-to-body bond that consists in two ionic bonds. In water the bonding energy between two ions is typically \SI{5}{\kilo\joule\per\mole}\cite{Schneider1992}, thus $E_c\approx 2 k_\mathrm{B}T$. However the medium surrounding the ionic bond cannot in general be described as pure water. As the chain is in poor solvent, electrostatic blobs have a low water content. Such a low polarity microenvironement is well known to enhance otherwise weak electrostatic interactions in protein folding or engineered self-assembly\cite{Rehm2010}.

From Equation~(\ref{eq:geDfromB}) we know that $D$ increases with $B$. We thus expect stronger head-to-body bonds when $B$ is large and the low-polarity environment is extended ($D\gg\ell_\mathrm{B}$). A contrario, the weakest ionic links, closest to their strength in water, should be found in \ce{PPyr+I-} where $B$ is minimum ($D\approx\ell_\mathrm{B}$).

Applying Equation~(\ref{eq:tau}) to \ce{PPyr+I-} with $E_c = 2 k_\mathrm{B}T$ yields $\tau \approx 0.40\pm 0.07$, or a $\Theta$-temperature around \SI{220}{\celsius}. Using this value and Equation (\ref{eq:stretch}) we deduce the bonding energy for each composition, as reported in Table~\ref{tab:results}. Indeed, the bonding energy increases with increasing counterion condensation to reach, in the case of \ce{PIm+Br-}, 85\% of the carbon-carbon single bond (\SI{350}{\kilo\joule\per\mole} or $140 k_\mathrm{B}T$). According to Equation (\ref{eq:geDfromB}), in this extreme case an electrostatic blob contains $\approx 10^3$ polymers which embed the head-to-body bonds in a low-polarity environment $\approx 24$ times larger than the Bjerrum length. Conversely, for \ce{PPyr+I-} and \ce{PPyr+Br-} the size of electrostatic blobs is comparable to $\ell_\mathrm{B}$, confirming an aqueous microenvironment. Quantitatively, the bonding energy of an ionic bond being inversely proportional to the relative dielectric constant $\epsilon_r$, we need to suppose a local $\epsilon_r$ around 2 (typical of water-insoluble polymers) instead of 80 for water to recover the large bonding energy of \ce{PIm+Br-}.

\subsubsection{Quantitative phase diagram.}

\begin{figure}
\includegraphics{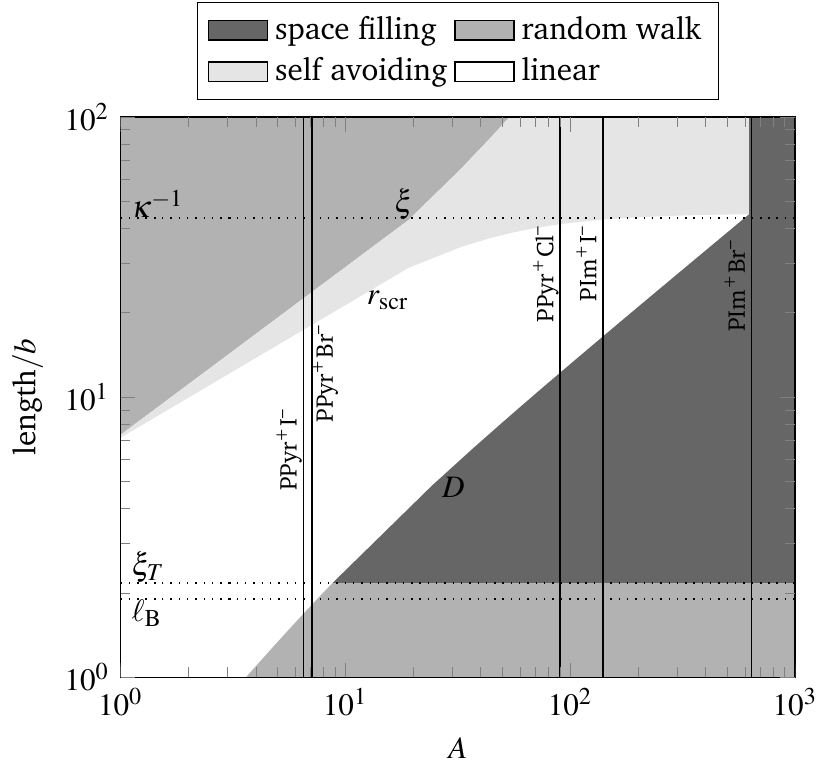}
\caption{Conformation of semi-dilute polyelectrolytes at rest at various scales as a function of counterion condensation parameter $A$. The phase diagram is given for a reduced temperature $\tau=0.46$ and for the monomer number density of \ce{PIm+Br-}, although $c$ is slightly different for each sample. Each vertical line represents a sample.}
\label{fig:phasediag}
\end{figure}

Table~\ref{tab:results} also shows the values of $A$ deduced from Equation~(\ref{eq:A}) and the measured value of $\tau$. $A$ varies between $7$ for \ce{PPyr+I-}, corresponding to $10$ uncondensed counterions per polymer, to $640$ for \ce{PIm+Br-} indicating a single dissociated counterion every $9$ polymers. We stress that even in this case of very strong counterion condensation an electrostatic blob still contains several ($\approx 90$) charges ensuring electrostatic repulsion and solubility,

Figure~\ref{fig:phasediag} summarizes the different conformations of a procession as a function of scale and counterion condensation for our experimental $c$ and $\tau$. The most striking feature of this phase diagram is the narrowing of the regime between $D$ and $r_\mathrm{scr}$ (white region in Figure~\ref{fig:phasediag}) as counterion condensation increases. As the size of the electrostatic blob crosses the screening length, electrostatic repulsion is no longer able to sustain solubility. This abrupt transition between an uniform gel and a precipitate is what separates our five soluble samples from insoluble \ce{PPyr+F-}, \ce{PIm+Cl-} and \ce{PIm+F-}. Our rheological measurements probe the structure of the procession from the Kuhn length to the distance between crosslinks that is approximately $r_\mathrm{scr}$ (upper limit of the white region in Figure~\ref{fig:phasediag}). \ce{PPyr+Cl-} and \ce{PIm+I-} display the typical structure of weakly charged polyelectrolytes in poor solvent. \ce{PPyr+I-} and \ce{PPyr+Br-} lie right in the crossover to strongly charged regime. Their correct description is possible only through Equation~(\ref{eq:geDfromB}). Finally \ce{PIm+Br-} is extremely close to the transition to insolubility (within experimental errors) which explains its extreme softness and wide plastic regime.

From purely rheological measurements, we have thus obtained microscopic information on the state of the gel: amount of charge condensation, head-to-body bonding energy and size of a procession.

\section{Conclusion}
With this work we demonstrated the versatility of our ``head-to-body'' electrostatic approach in the fabrication of hydrogels with readily tunable rheological properties. Thus, depending on the nature of the nucleophile used in the quaternarization step of the polymer, but also on that of the counterion which can be modified in the course of its purification process, we showed that it is possible to manipulate almost at will the gel formability, the density of physical cross-links and the respective magnitudes of the shear modulus and of the shear stress. 

Strongly interacting anions (i.e. small and hard halides, like fluoride) and aromatic cations favour counterion condensation, resulting in too few charges to allow dissolution and gel formation. The hardest gels with the narrowest linear domain are obtained with aliphatic cationic moiety and counterion at the other end of the Hofmeister series that are less bound to the polymer. This gives rise to minimal counterion condensation, free the peripheral, iterative cations and increase their pairing probability with the terminal anion and thus affords a very high density of cross-links. In between, in the case of aliphatic cations with a counterion in the middle of the Hofmeister series (\ce{Cl-}) or in the case of aromatic cations and poorly interacting anion, counterion condensation is important and polymers associate in long processions with strong head-to-body bonds. We thus obtain soft gels able to sustain large deformations before flowing. Coincidentally, the mesh size of our gels is always close to the procession persistence length, a regime often encountered in networks of biological semiflexible filaments as collagen or actin\cite{Meng2016}.

To conclude, our procedure yields robust, highly tunable hydrogels from short, linear polymer chains and in the absence of any additive, which could find interesting applications especially in the context of biomaterials. Most importantly, systematic comparisons between the different poly(electrolytes) investigated in this study were used to establish a model linking the microstructure of the gel to ion pair dissociation efficiency on the individual polymer chains. We hope that this model will join the procession toward future works in supramolecular assemblies.

\paragraph*{Acknowledgements.}
The authors are grateful for the financial support of Universit\'{e} de Lyon through the program ``Investissements d'Avenir'' (ANR-1 1-IDEX-0007). S.M. and N.T. acknowledge funding from the European Research Council under the European Union's Seventh Framework Program (FP7/2007-2013)/ERC grant agreement No.258803. The Graphical TOC entry is derived from a picture by Arturo Reina, CC BY-SA 3.0, \url{https://commons.wikimedia.org/w/index.php?curid=282898}.

\bibliography{../PALSE} 

\end{document}